\documentclass[%
 reprint,
superscriptaddress,
 amsmath,amssymb,
 aps,
]{revtex4-2}
\usepackage[version=4]{mhchem}
\usepackage{soul}
\usepackage{siunitx}
\usepackage{color}
\usepackage{graphicx}
\usepackage{dcolumn}
\usepackage{bm}
\usepackage{hyperref}
\hypersetup{colorlinks=true,linkcolor=red,citecolor=blue}

\begin{document}

\preprint{APS/123-QED}
\title{The Double-$Q$ Ground State with Topological Charge Stripes in the Skyrmion Candidate \ce{GdRu2Si2}}

\author{G. D. A. Wood}
\email{George.Wood@warwick.ac.uk}
\affiliation{Department of Physics, University of Warwick, Coventry, CV4 7AL, United Kingdom}
\author{D. D. Khalyavin}
\affiliation{ISIS Facility, STFC Rutherford Appleton Laboratory, Harwell Science and Innovation Campus, Oxfordshire OX11 0QX, United Kingdom}
\author{D. A. Mayoh}
\affiliation{Department of Physics, University of Warwick, Coventry, CV4 7AL, United Kingdom}
\author{J. Bouaziz}
\affiliation{Peter Gr\"unberg Institut and Institute for Advanced Simulation, Forschungszentrum J\"ulich \& JARA, D-52425 J\"ulich, Germany}
\author{A. E. Hall}
\affiliation{Department of Physics, University of Warwick, Coventry, CV4 7AL, United Kingdom}
\author{S. J. R. Holt}
\affiliation{Faculty of Engineering and Physical Sciences, University of Southampton, Southampton SO17 1BJ, United Kingdom}
\affiliation{Max Planck Institute for the Structure and Dynamics of Matter,
Luruper Chaussee 149, 22761 Hamburg, Germany}
\author{F. Orlandi}
\affiliation{ISIS Facility, STFC Rutherford Appleton Laboratory, Harwell Science and Innovation Campus, Oxfordshire OX11 0QX, United Kingdom}
\author{P. Manuel}
\affiliation{ISIS Facility, STFC Rutherford Appleton Laboratory, Harwell Science and Innovation Campus, Oxfordshire OX11 0QX, United Kingdom}
\author{S. Bl\"ugel}
\affiliation{Peter Gr\"unberg Institut and Institute for Advanced Simulation, Forschungszentrum J\"ulich \& JARA, D-52425 J\"ulich, Germany}
\author{J. B. Staunton}
\affiliation{Department of Physics, University of Warwick, Coventry, CV4 7AL, United Kingdom}
\author{O. A. Petrenko}
\affiliation{Department of Physics, University of Warwick, Coventry, CV4 7AL, United Kingdom}
\author{M. R. Lees}
\affiliation{Department of Physics, University of Warwick, Coventry, CV4 7AL, United Kingdom}
\author{G. Balakrishnan}
\email{G.Balakrishnan@warwick.ac.uk}
\affiliation{Department of Physics, University of Warwick, Coventry, CV4 7AL, United Kingdom}

\date{\today}

\begin{abstract}


\ce{GdRu2Si2} is a centrosymmetric magnet in which a skyrmion lattice has recently been discovered. Here, we investigate the magnetic structure of the zero field ground state using neutron diffraction on single crystal and polycrystalline \ce{^{160}GdRu2Si2}. In addition to observing the principal propagation vectors $\mathbf{q}_{1}$ and $\mathbf{q}_{2}$, we discover higher order magnetic satellites, notably $\mathbf{q}_{1} + 2\mathbf{q}_{2}$. The appearance of these satellites are explained within the framework of a new double-$Q$ constant-moment solution. Using powder diffraction we implement a quantitative refinement of this model. This structure, which contains vortexlike motifs, is shown to have a one-dimensional topological charge density. 
\end{abstract}

\maketitle



Skyrmions and merons are two-dimensional (2D) vortexlike spin textures which have integer and half-integer topological charge respectively~\cite{lancaster2019skyrmions}. These non-collinear spin textures exhibit a range of emergent phenomena such as the topological Hall effect and these properties are anticipated to find use in low-power spintronics~\cite{neubauer2009topological,fert2017magnetic}. Magnetic skyrmion research has principally been focused on noncentrosymmetric materials such as \ce{MnSi} and \ce{Cu2OSeO3}, where the Dzyaloshinskii-Moriya interaction plays a key role in the stabilization of skyrmions in a magnetic field~\cite{sm2009skyrmion,seki2012observation}. 

In contrast, the interactions stabilizing skyrmions in centrosymmetric materials are varied and in general less well understood. Notably though, experiments have determined the formation of skyrmions in a set of centrosymmetric \ce{Gd}/\ce{Eu}-based intermetallics: \ce{Gd2PdSi3}~\cite{kurumaji2019skyrmion}, \ce{Gd3Ru4Al12}~\cite{hirschberger2019skyrmion}, \ce{EuAl4}~\cite{takagi2022square}, \ce{EuGa2Al2}~\cite{PhysRevMaterials.6.074201}, and of focus to this letter \ce{GdRu2Si2}~\cite{khanh2020nanometric, yasui2020imaging, khanh2022zoology}. \ce{Eu^{2+}} and \ce{Gd^{3+}} ions have long been established as prototypical Heisenberg spin systems ($S=7/2,L=0$), these ions are therefore an important inclusion in this class of materials, although not essential~\cite{hou2021emergence}. In \ce{Gd2PdSi3} and \ce{Gd3Ru4Al12} geometrical frustration is expected to be a key consideration for the formation of the skyrmion lattices, whereas for \ce{GdRu2Si2}, \ce{EuAl4} and \ce{EuGa2Al2}, which crystallize in the $I4/mmm$ space group, this cannot be the case.

\begin{figure}[tb]
\includegraphics*[width=1\linewidth,clip]{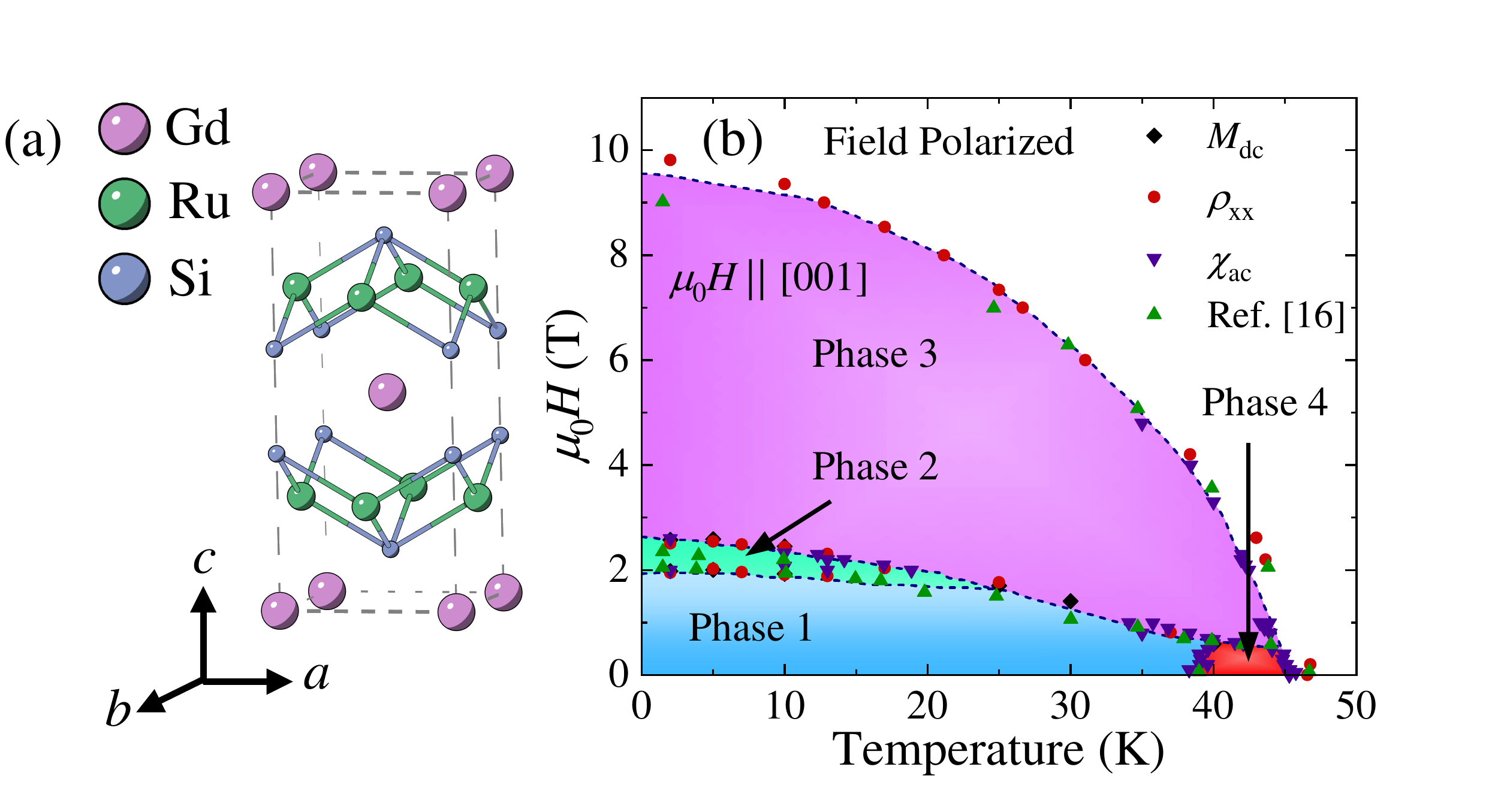}
\caption{\label{fig:HTPD} (a) Crystal structure of \ce{GdRu2Si2}, space group: $I4/mmm$. (b) \textit{H-T} phase diagram of \ce{GdRu2Si2} with $H\parallel c$. Phase boundaries have been established using a combination of longitudinal resistivity ($\rho_{\mathrm{xx}}$), DC magnetometry ($M_{\mathrm{dc}}$) and AC susceptibility ($\chi_{\mathrm{ac}}$). This phase diagram is consistent with previous determinations~\cite{garnier1996giant}. Phases 1 and 3 have been nominally designated as meron lattices and Phase 2 the skyrmion lattice, with no current interpretation of Phase 4~\cite{khanh2022zoology}.}
\end{figure}

For a consistent theory of the magnetism in \ce{GdRu2Si2}, it is necessary to establish the magnetic structures across the phase diagram shown in Fig.~\ref{fig:HTPD}. Resonant elastic x-ray scattering (REXS) studies reported two incommensurate propagation vectors, denoted as $\mathbf{q}_{1}$ and $\mathbf{q}_{2}$ along the \textit{a} and \textit{b} axes respectively~\cite{khanh2020nanometric,khanh2022zoology}. Initially, Phase 1 was suggested to be a single-$Q$ helix state and Phase 2 as the square skyrmion lattice, composed of two superimposed orthogonal helices~\cite{khanh2020nanometric}. An \textit{ab initio} study predicted that the incommensurate helical state was due to Fermi surface nesting, reproducing propagation vectors and a critical temperature ($T_{\mathrm{c}}$) in close accordance with experiments. Moreover, that study predicted that the skyrmion lattice could be formed with just first-order Ruderman–Kittel–Kasuya–Yosida (RKKY) interactions, magnetic anisotropy and Zeeman coupling~\cite{bouaziz2022fermi}. Subsequently, a spectroscopic scanning tunneling microscopy experiment identified double-$Q$ incommensurate ordering in Phase 1, detecting $\mathbf{q}_{1}+\mathbf{q}_{2}$ and $2\mathbf{q}_{1,2}$ distortions~\cite{yasui2020imaging}. The identification of incommensurate electron density coincident with the magnetic order strengthens arguments for RKKY exchange~\cite{bouaziz2022fermi}, as opposed to alternative mechanisms which had previously been suggested~\cite{PhysRevLett.125.117204}. Following the indications of 2D ordering across the phase diagram, phases 1 and 3 were assigned as meron lattices with Phase 1 constructed from a helix on $\mathbf{q}_{1}$ and a spin-density wave (SDW) along $\mathbf{q}_{2}$. To stabilize this ground state, biquadratic exchange terms are required as shown in the REXS/Monte Carlo annealing study~\cite{khanh2022zoology}. No microscopic description has been given for Phase 4, however, the phase boundaries are detectable using magnetometry. The lack of information regarding this phase is probably due to the magnetism being weak with increased disorder in proximity to $T_{\mathrm{c}}$ ($45$~K) and is a topic of further study.

In this Letter, the zero field ground state of \ce{GdRu2Si2} is examined using time-of-flight neutron diffraction. Within the $\left(h,k,0\right)$ diffraction plane, magnetic satellites of the form $\mathbf{q}_{1}$, $\mathbf{q}_{2}$ and $\mathbf{q}_{1}+2\mathbf{q}_{2}$ are found. Furthermore, the principal propagation vectors $\mathbf{q}_{1}$ and $\mathbf{q}_{2}$ are shown to be inequivalent in magnitude. These aspects taken together, allow the analytical construction of a double-$Q$ constant-moment solution, which is entropically favorable. This model is used in the quantitative refinement of the powder diffraction, giving a complete magnetic structure solution. The topological charge density (TCD) of this phase is particularized and shown to have one-dimensional (1D) topological charge stripes. Finally the implications of this new magnetic structure on the Hamiltonian are discussed.



To reduce neutron absorption isotopically enriched $^{160}$Gd was used. Polycrystalline $^{160}$GdRu$_{2}$Si$_{2}$ was prepared by arc melting stoichiometric quantities of the constituent materials. A single crystal of isotopically-enriched $^{160}$GdRu$_{2}$Si$_{2}$, 17~mm in length and 2.3~mm in diameter was grown by the floating zone method.

Polycrystalline and single crystal neutron diffraction measurements were taken on the WISH diffractometer at the ISIS Neutron and Muon Source UK~\cite{chapon2011wish}. The measurements were taken in Phase 1 in zero magnetic field and at $1.5~\si{K}$ to maximize the magnetic scattering signal. Polycrystalline $^{160}$GdRu$_{2}$Si$_{2}$ was ground into a fine powder and placed in a vanadium can. The single crystal was aligned so that $\left(h,k,0\right)$ was the principal scattering plane.

\begin{figure}[b]
\includegraphics*[width=1\linewidth,clip]{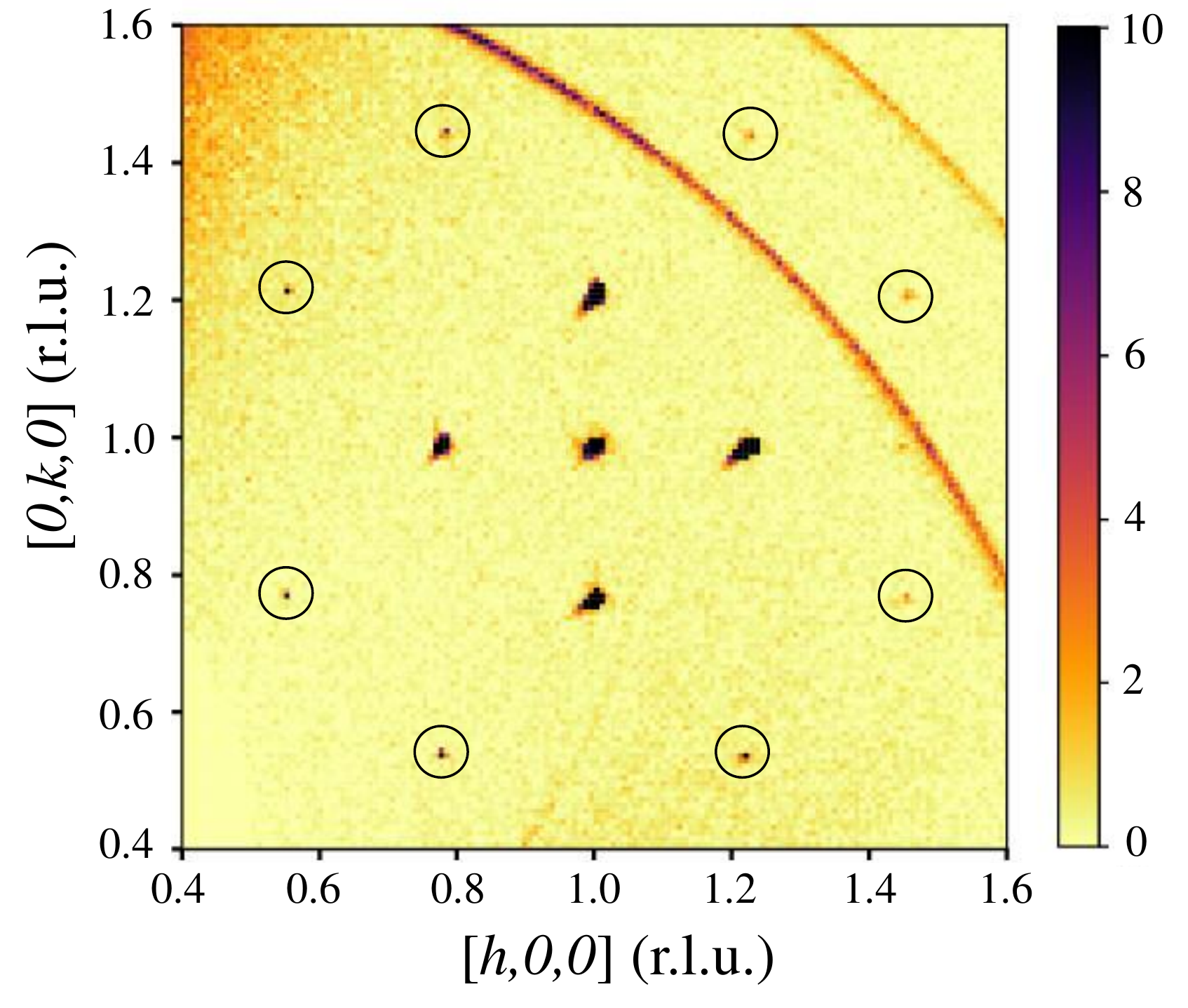}
\caption{\label{fig:HKSC} Single crystal neutron diffraction data in the $(h,k,0)$ plane showing magnetic satellites surrounding the $\left(1,1,0\right)$ nuclear peak at $1.5~\si{K}$. Intensity has been scaled so that the nuclear peak and  sets of incommensurate satellites of the form: $\mathbf{q}_{1}$, $\mathbf{q}_{2}$ and $\mathbf{q}_{1}+2\mathbf{q}_{2}$ (circled in black), can all be simultaneously observed. The arcs are  \ce{Al} powder diffraction lines from the sample holder.}
\end{figure}

\begin{figure}[t]
\includegraphics*[width=1\linewidth,clip]{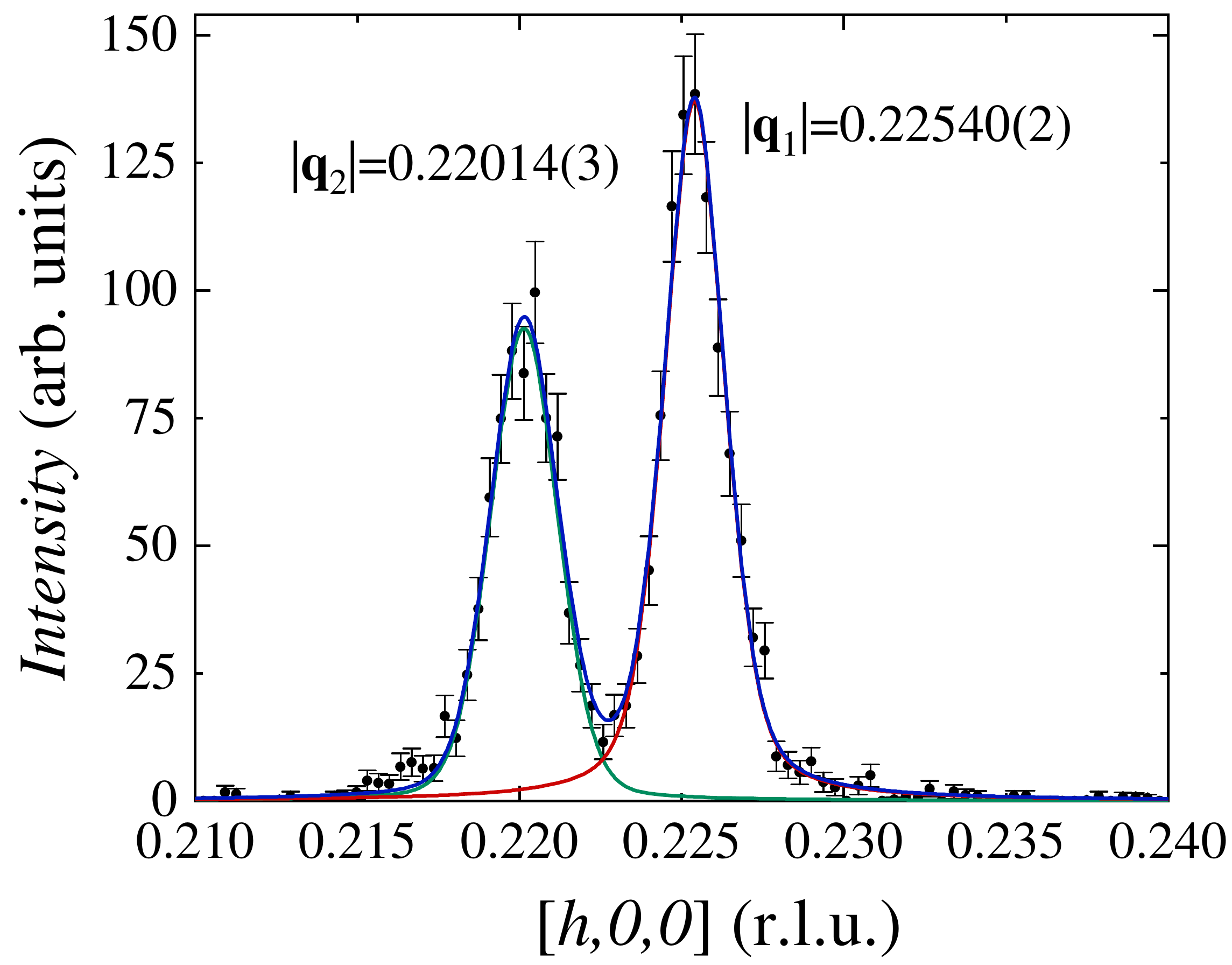}
\caption{\label{fig:q1q2} Single crystal neutron diffraction data showing the splitting between inequivalent $\mathbf{q}_{1}$ and $\mathbf{q}_{2}$ at $1.5~\si{K}$. The peaks have been fit with a pseudo-Voigt function. These measurements are of the $\left(0+q_{1},0,0\right)$ and $\left(0+q_{2},0,0\right)$ satellites.}
\end{figure}

 Magnetic satellites surrounding the $\left(1,1,0\right)$  nuclear peak are shown in Fig ~\ref{fig:HKSC}. Attention was given to this peak, in part as it was convenient to access the sets of magnetic satellites, and also the small scattering vector ensured that the magnetic form factor was large. The magnetic diffraction shown is due to two orthogonal magnetic domains, however this is only clear in subsequent measurements (Fig. \ref{fig:q1q2}) which resolve the inequivalence between the $\mathbf{q}_{1}$ and $\mathbf{q}_{2}$ propagation vectors. The two domains are denoted with $A$ and $B$ subscripts respectively. The set of satellites observed is $\left(1\pm q_{1},1,0\right)_{A}$, $\left(1,1 \pm q_{2},0\right)_{A}$, $\left(1\pm q_{1},1 \pm 2q_{2},0\right)_{A}$, $\left(1,1 \pm q_{1},0\right)_{B}$, $\left(1\pm q_{2},1,0\right)_{B}$ and $\left(1\pm 2q_{2},1 \pm q_{1},0\right)_{B}$ . See Fig. S1 for a schematic of how each magnetic domain contributes to the diffraction plane \cite{SI}. Whilst the existence of $\mathbf{q}_{1}$ and $\mathbf{q}_{2}$ satellites had been well established this is the first instance in which $\mathbf{q}_{1}+2\mathbf{q}_{2}$ satellites have been observed in a diffraction experiment.  We note that surrounding other nuclear peaks, satellites of the form $\mathbf{q}_{1}+\mathbf{q}_{2}$ were infrequently observed (see Fig. S2 for an example \cite{SI}). The presence of $\mathbf{q}_{1}+\mathbf{q}_{2}$ and $\mathbf{q}_{1}+2\mathbf{q}_{2}$ satellites are independent proof of the double-$Q$ magnetic structure, coming from the evaluation of the possible free-energy coupling terms~\cite{izyumov1990phase}. The associated distortions ($\mathbf{q}_{1}+\mathbf{q}_{2}$ and $\mathbf{q}_{1}+2\mathbf{q}_{2}$) are produced by the two principal arms of the star ($\mathbf{q}_{1}$ and $\mathbf{q}_{2}$) being coupled together in the ground state magnetic structure. Moreover, the translation and time reversal symmetries require a third power invariant for the $\mathbf{q}_{1}+\mathbf{q}_{2}$ and a fourth power invariant for the $\mathbf{q}_{1}+2\mathbf{q}_{2}$ distortions. This in turn implies that the former is of structural origin and the latter is magnetic.


Additionally, the existence of these higher-order satellites indicates that the magnetic structure may have a constant-moment solution, whereas the magnetic structure solutions described in this material hitherto have varying spin-density.  Strongly motivating this notion, the saturation magnetization of \ce{GdRu2Si2} has been shown to be $7~\mu_{\mathrm{B}}$~\cite{garnier1996giant, khanh2020nanometric}. This substantiates that the magnetism is formed of the localized \textit{f}-electron moments of \ce{Gd^{3+}} which interact via valence electrons at the Fermi surface. This model is inconsistent with a system of varying spin-density, as this would incur a significant entropic cost due to disorder and increased kinetic energy, which is not expected in the zero field ground state. Indeed, established SDW systems ordinarily have particular reasons for how the SDW is stabilized. In itinerant magnets this is due to wave-vector nesting in the Fermi surface~\cite{fawcett1988spin}, or in insulators frustrated exchange interactions can generate these unusual states~\cite{PhysRevLett.101.207201, PhysRevB.87.054408, PhysRevLett.129.087201}. However, \ce{GdRu2Si2} is in neither of these categories of materials.

 Before the constant-moment solution can be constructed, it is necessary to establish the inequivalence in magnitude between $\mathbf{q}_{1}$ and $\mathbf{q}_{2}$~\cite{khanh2022zoology}. This is shown in Fig.~\ref{fig:q1q2}, which shows measurements of satellites surrounding the Brillouin zone center: $\left(0+q_{1},0,0\right)$ and $\left(0+q_{2},0,0\right)$. These satellites are observed in the same region of reciprocal space due to the ground state forming in two mutually orthogonal domains, reducing the symmetry of the parent tetragonal structure. By the Curie principle, this induces (so far undetected) nuclear distortions with identical domain formation.
 


Having established the $\mathbf{q}_{1}+2\mathbf{q}_{2}$ satellites and the inequivalence between $\mathbf{q}_{1}$ and $\mathbf{q}_{2}$, the constant-moment solution can be fully described. The ground state of \ce{GdRu2Si2} has been previously proposed to comprise a helix and an orthogonal SDW propagating along $\mathbf{q}_{1}$ and $\mathbf{q}_{2}$, respectively~\cite{khanh2022zoology}. The construction of a constant-moment solution includes the same helix and SDW with the addition of a second helix, coplanar with the first, but propagating along $\mathbf{q}_{1}+2\mathbf{q}_{2}$ as an ansatz. The magnetic moment $\mathbf{m_{r}}$, at a position vector $\mathbf{r}$ is then given by,
\begin{equation}
    \mathbf{m_{r}}=\begin{bmatrix} 0 \\m_{1}\cos(k_{1})\\m_{1}\sin(k_{1})\end{bmatrix} + \begin{bmatrix}m_{2}\sin(k_{2}) \\0 \\0 \end{bmatrix} + \begin{bmatrix}0 \\m_{3}\cos(k_{1}+2k_{2}) \\m_{3}\sin(k_{1}+2k_{2})\end{bmatrix}.
\label{solution}
\end{equation}

\noindent Here, $k_{1,2}=-2\pi\boldsymbol{q_{1,2}\cdot r}$ and $m_{1,2,3}$ are scalar amplitudes. Note that Eq. \ref{solution} holds for any choice of two propagation vectors, and that these planar waves can be defined on an arbitrary orthonormal basis. For a constant-moment solution $\mathbf{m_{r}}\cdot\mathbf{m_{r}}$ is evaluated,

\begin{equation}
\mathbf{m_{r}}\cdot\mathbf{m_{r}}=m_{1}^{2} + m_{3}^{2} + 4m_{1}m_{3}\cos^{2}(k_{2})-2m_{1}m_{3}+ m_{2}^{2}\sin^{2}(k_{2}),
\label{mag}
\end{equation}
\noindent with the condition $m_{3} = m_{2}^{2}/4m_{1}$. Hence a constant-moment solution can be provided in the form of Eq.~\ref{solution} with an appropriate moment condition.

\begin{figure}[t]
\includegraphics*[width=1\linewidth,clip]{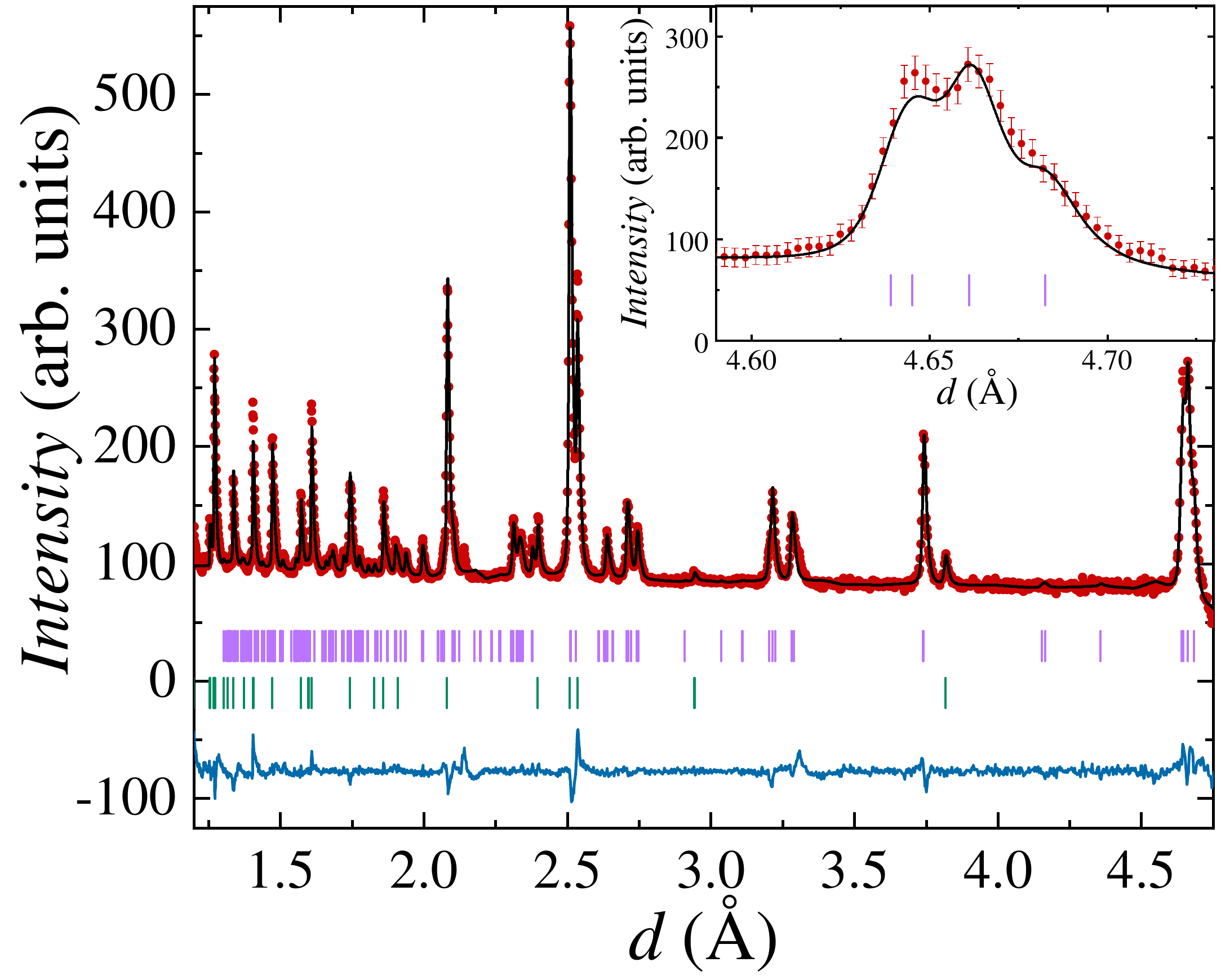}
\caption{\label{fig:PR} Rietveld refinement of \ce{GdRu2Si2} at $1.5~\si{K}$. Red points are the data, the black curve is the model and the blue line is the difference curve. Green and purple bars indicate the nuclear and magnetic Bragg peaks, respectively. This pattern is from the high-resolution bank of the WISH diffractometer, where $\mathbf{q}_{1}$ and $\mathbf{q}_{2}$ can be resolved, as highlighted in the inset. $R_{wp}=15.3\%$ and $R_{exp}=8.73\%$ and $GOF=1.75$.}
\end{figure}

\begin{figure}[t]
\includegraphics*[width=1\linewidth,clip]{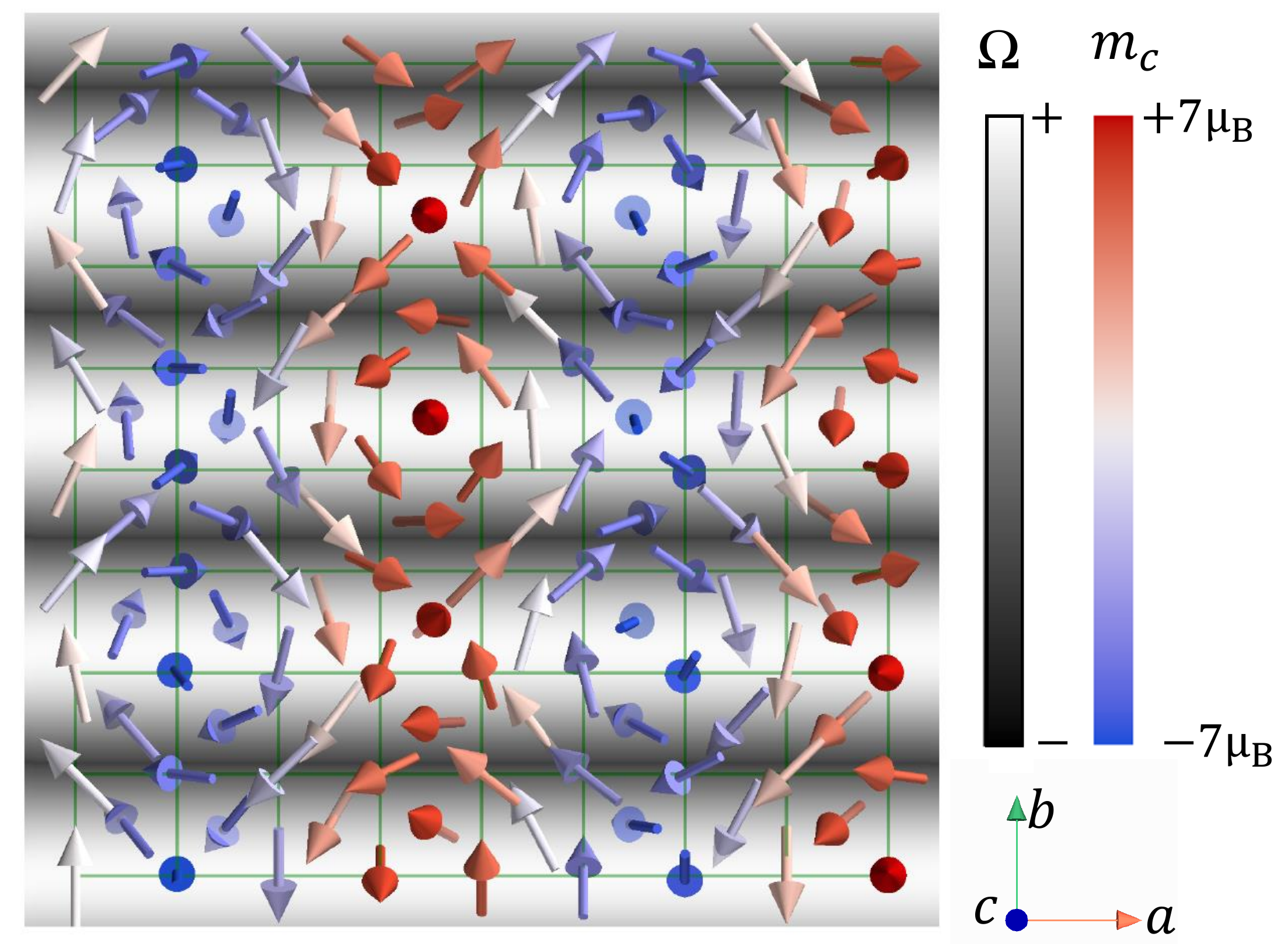}
\caption{\label{fig:MS} A model of the magnetic structure, shown over $8\times8$ unit cells. The out-of-plane component of the magnetic moment $m_{c}$ is shown with the blue-red color scale. The topological charge density is represented with the black-white color scale, showing the topological charge stripes. }
\end{figure}

Quantitative evaluation of the above model was implemented by analyzing the powder neutron diffraction (Fig. ~\ref{fig:PR}) using \textsc{fullprof}~\cite{rodriguez1993recent}. The most appropriate choice of basis was found to be $\mathbf{\hat{a}}=[1,0,0]$,~$\mathbf{\hat{b}}=[0,1,0]$, and ~$\mathbf{\hat{c}}=[0,0,1]$, with $\mathbf{q}_{1} = [0.22540(2),0,0]$ and $\mathbf{q}_{2} = [0,0.22014(3),0]$. The scalar magnitudes were found to be: $m_{1}=(6.49\pm0.10) ~\mu_{\mathrm{B}}$, $m_{2}=(5.44\pm0.16)~\mu_{\mathrm{B}}$ and $m_{3}=(1.14\pm0.09)~\mu_{\mathrm{B}}$. Note that in the powder data there were insufficient statistics to independently fit the $\mathbf{q}_{1}+2\mathbf{q}_{2}$ satellite peaks, and therefore the value of $m_{3}$ is implicit through the derived moment condition.  From Eq. \ref{mag} this gives a total moment size of $\left | \mathbf{m_{r}}  \right |=(7.6\pm0.1)~\mu_{\mathrm{B}}$. The discrepancy between the measured moment size and the $7~\mu_{\mathrm{B}}$ of \ce{Gd^{3+}} is caused by a systematic error, attributed to the high levels of absorption in the sample, which is heavily correlated to the scale factor and in turn the refined moment size. This magnetic structure is illustrated in Fig.~\ref{fig:MS}, where periodic vortexlike motifs are observed. 
 

The topological charge structure is now examined. Our analysis is based in the continuous field limit, but we note that in the more discrete limit of a small unit cell such as in this system, the notion of TCD is less clear. TCD is given by,
\begin{equation}
\Omega = \mathbf{\hat{m_{r}}} \cdot \frac{\partial \mathbf{\hat{m_{r}}} }{\partial x} \times \frac{\partial \mathbf{\hat{m_{r}}} }{\partial y},
\label{TCD}
\end{equation}
\noindent where $\mathbf{\hat{m_{r}}} = \mathbf{m_{r}}/\left| \mathbf{m_{r}}\right|$. Substituting Eq.~\ref{solution} into Eq.~\ref{TCD} yields,
\begin{equation}
\Omega =  \frac{(2\pi)^{2}4q_{1}q_{2}m_{1}m_{2}}{4m_{1}^{2} + m_{2}^{2}} \cos(-2\pi q_{2}y).
\label{TCDM}
\end{equation}
 \noindent Here the Cartesian axes are associated with the $\mathbf{\hat{a}}$, $\mathbf{\hat{b}}$ and $\mathbf{\hat{c}}$ directions. Hence the TCD is locally non-trivial, 1D, and oscillates in the same direction as the SDW component of the magnetic structure. We refer to this defining feature as topological charge stripes. Additionally, this establishes that the spin texture of this ground state cannot be compatible with a periodic meron or skyrmion lattice. This is since these spin textures would have a 2D periodic TCD structure, over which a 2D integral to evaluate the topological charge can be well defined with periodic boundary conditions. With the TCD structure given in Eq.~\ref{TCDM}, a 2D integral to evaluate the topological charge over this scalar field would be arbitrary given there is only periodicity in a single direction. We expect that these topological charge stripes may produce novel transport phenomena, in which there may be significant anisotropy perpendicular and parallel to the propagation of the TCD due to the topological Hall effect. 

A first principles analysis of the Fourier transform of the magnetic 
interactions in \ce{GdRu2Si2} assuming fourfold symmetry (without magneto-striction effects) reveals the presence of peaks at four 
characteristic wave vectors $\mathbf{q}_{1}=(\pm q_\text{max},0,0)$ and $\mathbf{q}_{2}=(0,\pm q_\text{max},0)$~\cite{bouaziz2022fermi}. This indicates the possibility of stabilizing single-$Q$ states at multiple wave vectors. The inclusion of multiple-spin interactions leads to the stabilization of multiple-$Q$ states~\cite{hayami2021topological}. Particularly, the biquadratic interactions play an important role in the 
stabilization of anisotropic double-$Q$ states~\cite{khanh2022zoology}. We postulate that the biquadratic interactions together with other more complex isotropic multiple-spin interactions, such as three-site four spin interactions~\cite{Hoffmann2020} and ring exchange~\cite{heinze2011spontaneous} can generate the complex magnetic structure reported here~\cite{sharma2022machine}. These higher-order spin interactions derive from the feedback effect on the valence electrons from the development of the complex magnetic order of the \ce{Gd} local moments~\cite{PhysRevLett.118.197202}.

In conclusion, the zero field ground state of \ce{GdRu2Si2} has been studied using time-of-flight neutron diffraction. In addition to observing the principal magnetic satellites $\mathbf{q}_{1}$ and $\mathbf{q}_{2}$ we discover satellites of the form $\mathbf{q}_{1}+2\mathbf{q}_{2}$. These magnetic satellites allow the analytical construction and quantitative refinement of a constant-moment solution, which is entropically favorable. The principles upon which the constant-moment solution is constructed are expected to be a novel aid in determining the complex nature of multi-$Q$ magnetic structures. The TCD of this magnetic structure is shown to be locally non-trivial, 1D, and it is not consistent with either meron or skyrmion lattices. We speculate that these topological charge stripes may give rise to novel magneto-electronic coupling, such as producing electronic nematicity. These results establish that \ce{GdRu2Si2} has a wealth of topologically non-trivial spin textures and is therefore an ideal setting in which phase transitions between distinct topological structures can be experimentally probed. An interpretation of the in-field phases and particularly Phase 4 remains a topic of further study.
\begin{acknowledgments}
We would like to acknowledge Tom Orton and Patrick Ruddy for their technical support. This Letter was financially supported by two Engineering and Physical Sciences Research Council grants: EP/T005963/1, and the UK Skyrmion Project Grant, EP/N032128/1. This Letter was supported by U.K. Research and Innovation and Science and Technology Facilities Council through the provision of beam time at the ISIS Neutron and Muon Source~\cite{Gdata1}, as well as partial funding for the \ce{^{160}Gd} isotope used in the neutron experiment.
\end{acknowledgments}

\bibliography{Main.bib}

\begin{thebibliography}{30}%
\makeatletter
\providecommand \@ifxundefined [1]{%
 \@ifx{#1\undefined}
}%
\providecommand \@ifnum [1]{%
 \ifnum #1\expandafter \@firstoftwo
 \else \expandafter \@secondoftwo
 \fi
}%
\providecommand \@ifx [1]{%
 \ifx #1\expandafter \@firstoftwo
 \else \expandafter \@secondoftwo
 \fi
}%
\providecommand \natexlab [1]{#1}%
\providecommand \enquote  [1]{``#1''}%
\providecommand \bibnamefont  [1]{#1}%
\providecommand \bibfnamefont [1]{#1}%
\providecommand \citenamefont [1]{#1}%
\providecommand \href@noop [0]{\@secondoftwo}%
\providecommand \href [0]{\begingroup \@sanitize@url \@href}%
\providecommand \@href[1]{\@@startlink{#1}\@@href}%
\providecommand \@@href[1]{\endgroup#1\@@endlink}%
\providecommand \@sanitize@url [0]{\catcode `\\12\catcode `\$12\catcode
  `\&12\catcode `\#12\catcode `\^12\catcode `\_12\catcode `\%12\relax}%
\providecommand \@@startlink[1]{}%
\providecommand \@@endlink[0]{}%
\providecommand \url  [0]{\begingroup\@sanitize@url \@url }%
\providecommand \@url [1]{\endgroup\@href {#1}{\urlprefix }}%
\providecommand \urlprefix  [0]{URL }%
\providecommand \Eprint [0]{\href }%
\providecommand \doibase [0]{https://doi.org/}%
\providecommand \selectlanguage [0]{\@gobble}%
\providecommand \bibinfo  [0]{\@secondoftwo}%
\providecommand \bibfield  [0]{\@secondoftwo}%
\providecommand \translation [1]{[#1]}%
\providecommand \BibitemOpen [0]{}%
\providecommand \bibitemStop [0]{}%
\providecommand \bibitemNoStop [0]{.\EOS\space}%
\providecommand \EOS [0]{\spacefactor3000\relax}%
\providecommand \BibitemShut  [1]{\csname bibitem#1\endcsname}%
\let\auto@bib@innerbib\@empty
\bibitem [{\citenamefont {Lancaster}(2019)}]{lancaster2019skyrmions}%
  \BibitemOpen
  \bibfield  {author} {\bibinfo {author} {\bibfnamefont {T.}~\bibnamefont
  {Lancaster}},\ }\bibfield  {title} {\bibinfo {title} {Skyrmions in magnetic
  materials},\ }\href
  {https://doi.org/https://doi.org/10.1080/00107514.2019.1699352} {\bibfield
  {journal} {\bibinfo  {journal} {Contemp. Phys.}\ }\textbf {\bibinfo {volume}
  {60}},\ \bibinfo {pages} {246} (\bibinfo {year} {2019})}\BibitemShut
  {NoStop}%
\bibitem [{\citenamefont {Neubauer}\ \emph {et~al.}(2009)\citenamefont
  {Neubauer}, \citenamefont {Pfleiderer}, \citenamefont {Binz}, \citenamefont
  {Rosch}, \citenamefont {Ritz}, \citenamefont {Niklowitz},\ and\ \citenamefont
  {B{\"o}ni}}]{neubauer2009topological}%
  \BibitemOpen
  \bibfield  {author} {\bibinfo {author} {\bibfnamefont {A.}~\bibnamefont
  {Neubauer}}, \bibinfo {author} {\bibfnamefont {C.}~\bibnamefont
  {Pfleiderer}}, \bibinfo {author} {\bibfnamefont {B.}~\bibnamefont {Binz}},
  \bibinfo {author} {\bibfnamefont {A.}~\bibnamefont {Rosch}}, \bibinfo
  {author} {\bibfnamefont {R.}~\bibnamefont {Ritz}}, \bibinfo {author}
  {\bibfnamefont {P.}~\bibnamefont {Niklowitz}},\ and\ \bibinfo {author}
  {\bibfnamefont {P.}~\bibnamefont {B{\"o}ni}},\ }\bibfield  {title} {\bibinfo
  {title} {Topological {H}all effect in the \textit{{A}} phase of \ce{MnSi}},\
  }\href {https://doi.org/10.1103/PhysRevLett.102.186602} {\bibfield  {journal}
  {\bibinfo  {journal} {Phys. Rev. Lett.}\ }\textbf {\bibinfo {volume} {102}},\
  \bibinfo {pages} {186602} (\bibinfo {year} {2009})}\BibitemShut {NoStop}%
\bibitem [{\citenamefont {Fert}\ \emph {et~al.}(2017)\citenamefont {Fert},
  \citenamefont {Reyren},\ and\ \citenamefont {Cros}}]{fert2017magnetic}%
  \BibitemOpen
  \bibfield  {author} {\bibinfo {author} {\bibfnamefont {A.}~\bibnamefont
  {Fert}}, \bibinfo {author} {\bibfnamefont {N.}~\bibnamefont {Reyren}},\ and\
  \bibinfo {author} {\bibfnamefont {V.}~\bibnamefont {Cros}},\ }\bibfield
  {title} {\bibinfo {title} {Magnetic skyrmions: advances in physics and
  potential applications},\ }\href
  {https://doi.org/https://doi.org/10.1038/natrevmats.2017.31} {\bibfield
  {journal} {\bibinfo  {journal} {Nat. Rev. Mater.}\ }\textbf {\bibinfo
  {volume} {2}},\ \bibinfo {pages} {1} (\bibinfo {year} {2017})}\BibitemShut
  {NoStop}%
\bibitem [{\citenamefont {M{\"{u}}hlbauer}\ \emph {et~al.}(2009)\citenamefont
  {M{\"{u}}hlbauer}, \citenamefont {Binz}, \citenamefont {Jonietz},
  \citenamefont {Pfleiderer}, \citenamefont {Rosch}, \citenamefont {Neubauer},
  \citenamefont {Georgii},\ and\ \citenamefont {B{\"{o}}ni}}]{sm2009skyrmion}%
  \BibitemOpen
  \bibfield  {author} {\bibinfo {author} {\bibfnamefont {S.}~\bibnamefont
  {M{\"{u}}hlbauer}}, \bibinfo {author} {\bibfnamefont {B.}~\bibnamefont
  {Binz}}, \bibinfo {author} {\bibfnamefont {F.}~\bibnamefont {Jonietz}},
  \bibinfo {author} {\bibfnamefont {C.}~\bibnamefont {Pfleiderer}}, \bibinfo
  {author} {\bibfnamefont {A.}~\bibnamefont {Rosch}}, \bibinfo {author}
  {\bibfnamefont {A.}~\bibnamefont {Neubauer}}, \bibinfo {author}
  {\bibfnamefont {R.}~\bibnamefont {Georgii}},\ and\ \bibinfo {author}
  {\bibfnamefont {P.}~\bibnamefont {B{\"{o}}ni}},\ }\bibfield  {title}
  {\bibinfo {title} {Skyrmion lattice in a chiral magnet},\ }\href
  {https://doi.org/https://doi.org/10.1126/science.1166767} {\bibfield
  {journal} {\bibinfo  {journal} {Science}\ }\textbf {\bibinfo {volume}
  {323}},\ \bibinfo {pages} {915} (\bibinfo {year} {2009})}\BibitemShut
  {NoStop}%
\bibitem [{\citenamefont {Seki}\ \emph {et~al.}(2012)\citenamefont {Seki},
  \citenamefont {Yu}, \citenamefont {Ishiwata},\ and\ \citenamefont
  {Tokura}}]{seki2012observation}%
  \BibitemOpen
  \bibfield  {author} {\bibinfo {author} {\bibfnamefont {S.}~\bibnamefont
  {Seki}}, \bibinfo {author} {\bibfnamefont {X.}~\bibnamefont {Yu}}, \bibinfo
  {author} {\bibfnamefont {S.}~\bibnamefont {Ishiwata}},\ and\ \bibinfo
  {author} {\bibfnamefont {Y.}~\bibnamefont {Tokura}},\ }\bibfield  {title}
  {\bibinfo {title} {Observation of skyrmions in a multiferroic material},\
  }\href {https://doi.org/DOI: 10.1126/science.1214143} {\bibfield  {journal}
  {\bibinfo  {journal} {Science}\ }\textbf {\bibinfo {volume} {336}},\ \bibinfo
  {pages} {198} (\bibinfo {year} {2012})}\BibitemShut {NoStop}%
\bibitem [{\citenamefont {Kurumaji}\ \emph {et~al.}(2019)\citenamefont
  {Kurumaji}, \citenamefont {Nakajima}, \citenamefont {Hirschberger},
  \citenamefont {Kikkawa}, \citenamefont {Yamasaki}, \citenamefont {Sagayama},
  \citenamefont {Nakao}, \citenamefont {Taguchi}, \citenamefont {Arima},\ and\
  \citenamefont {Tokura}}]{kurumaji2019skyrmion}%
  \BibitemOpen
  \bibfield  {author} {\bibinfo {author} {\bibfnamefont {T.}~\bibnamefont
  {Kurumaji}}, \bibinfo {author} {\bibfnamefont {T.}~\bibnamefont {Nakajima}},
  \bibinfo {author} {\bibfnamefont {M.}~\bibnamefont {Hirschberger}}, \bibinfo
  {author} {\bibfnamefont {A.}~\bibnamefont {Kikkawa}}, \bibinfo {author}
  {\bibfnamefont {Y.}~\bibnamefont {Yamasaki}}, \bibinfo {author}
  {\bibfnamefont {H.}~\bibnamefont {Sagayama}}, \bibinfo {author}
  {\bibfnamefont {H.}~\bibnamefont {Nakao}}, \bibinfo {author} {\bibfnamefont
  {Y.}~\bibnamefont {Taguchi}}, \bibinfo {author} {\bibfnamefont {T.-H.}\
  \bibnamefont {Arima}},\ and\ \bibinfo {author} {\bibfnamefont
  {Y.}~\bibnamefont {Tokura}},\ }\bibfield  {title} {\bibinfo {title} {Skyrmion
  lattice with a giant topological hall effect in a frustrated
  triangular-lattice magnet},\ }\href
  {https://doi.org/https://doi.org/10.1126/science.aau0968} {\bibfield
  {journal} {\bibinfo  {journal} {Science}\ }\textbf {\bibinfo {volume}
  {365}},\ \bibinfo {pages} {914} (\bibinfo {year} {2019})}\BibitemShut
  {NoStop}%
\bibitem [{\citenamefont {Hirschberger}\ \emph {et~al.}(2019)\citenamefont
  {Hirschberger}, \citenamefont {Nakajima}, \citenamefont {Gao}, \citenamefont
  {Peng}, \citenamefont {Kikkawa}, \citenamefont {Kurumaji}, \citenamefont
  {Kriener}, \citenamefont {Yamasaki}, \citenamefont {Sagayama}, \citenamefont
  {Nakao} \emph {et~al.}}]{hirschberger2019skyrmion}%
  \BibitemOpen
  \bibfield  {author} {\bibinfo {author} {\bibfnamefont {M.}~\bibnamefont
  {Hirschberger}}, \bibinfo {author} {\bibfnamefont {T.}~\bibnamefont
  {Nakajima}}, \bibinfo {author} {\bibfnamefont {S.}~\bibnamefont {Gao}},
  \bibinfo {author} {\bibfnamefont {L.}~\bibnamefont {Peng}}, \bibinfo {author}
  {\bibfnamefont {A.}~\bibnamefont {Kikkawa}}, \bibinfo {author} {\bibfnamefont
  {T.}~\bibnamefont {Kurumaji}}, \bibinfo {author} {\bibfnamefont
  {M.}~\bibnamefont {Kriener}}, \bibinfo {author} {\bibfnamefont
  {Y.}~\bibnamefont {Yamasaki}}, \bibinfo {author} {\bibfnamefont
  {H.}~\bibnamefont {Sagayama}}, \bibinfo {author} {\bibfnamefont
  {H.}~\bibnamefont {Nakao}}, \emph {et~al.},\ }\bibfield  {title} {\bibinfo
  {title} {Skyrmion phase and competing magnetic orders on a breathing
  kagom{\'e} lattice},\ }\href
  {https://doi.org/https://doi.org/10.1038/s41467-019-13675-4} {\bibfield
  {journal} {\bibinfo  {journal} {Nat. Commun.}\ }\textbf {\bibinfo {volume}
  {10}},\ \bibinfo {pages} {5831} (\bibinfo {year} {2019})}\BibitemShut
  {NoStop}%
\bibitem [{\citenamefont {Takagi}\ \emph {et~al.}(2022)\citenamefont {Takagi},
  \citenamefont {Matsuyama}, \citenamefont {Ukleev}, \citenamefont {Yu},
  \citenamefont {White}, \citenamefont {Francoual}, \citenamefont {Mardegan},
  \citenamefont {Hayami}, \citenamefont {Saito}, \citenamefont {Kaneko} \emph
  {et~al.}}]{takagi2022square}%
  \BibitemOpen
  \bibfield  {author} {\bibinfo {author} {\bibfnamefont {R.}~\bibnamefont
  {Takagi}}, \bibinfo {author} {\bibfnamefont {N.}~\bibnamefont {Matsuyama}},
  \bibinfo {author} {\bibfnamefont {V.}~\bibnamefont {Ukleev}}, \bibinfo
  {author} {\bibfnamefont {L.}~\bibnamefont {Yu}}, \bibinfo {author}
  {\bibfnamefont {J.~S.}\ \bibnamefont {White}}, \bibinfo {author}
  {\bibfnamefont {S.}~\bibnamefont {Francoual}}, \bibinfo {author}
  {\bibfnamefont {J.~R.}\ \bibnamefont {Mardegan}}, \bibinfo {author}
  {\bibfnamefont {S.}~\bibnamefont {Hayami}}, \bibinfo {author} {\bibfnamefont
  {H.}~\bibnamefont {Saito}}, \bibinfo {author} {\bibfnamefont
  {K.}~\bibnamefont {Kaneko}}, \emph {et~al.},\ }\bibfield  {title} {\bibinfo
  {title} {Square and rhombic lattices of magnetic skyrmions in a
  centrosymmetric binary compound},\ }\href
  {https://doi.org/https://doi.org/10.1038/s41467-022-29131-9} {\bibfield
  {journal} {\bibinfo  {journal} {Nat. Commun.}\ }\textbf {\bibinfo {volume}
  {13}},\ \bibinfo {pages} {1} (\bibinfo {year} {2022})}\BibitemShut {NoStop}%
\bibitem [{\citenamefont {Moya}\ \emph {et~al.}(2022)\citenamefont {Moya},
  \citenamefont {Lei}, \citenamefont {Clements}, \citenamefont {Kengle},
  \citenamefont {Sun}, \citenamefont {Allen}, \citenamefont {Li}, \citenamefont
  {Peng}, \citenamefont {Husain}, \citenamefont {Mitrano}, \citenamefont
  {Krogstad}, \citenamefont {Osborn}, \citenamefont {Puthirath}, \citenamefont
  {Chi}, \citenamefont {Debeer-Schmitt}, \citenamefont {Gaudet}, \citenamefont
  {Abbamonte}, \citenamefont {Lynn},\ and\ \citenamefont
  {Morosan}}]{PhysRevMaterials.6.074201}%
  \BibitemOpen
  \bibfield  {author} {\bibinfo {author} {\bibfnamefont {J.~M.}\ \bibnamefont
  {Moya}}, \bibinfo {author} {\bibfnamefont {S.}~\bibnamefont {Lei}}, \bibinfo
  {author} {\bibfnamefont {E.~M.}\ \bibnamefont {Clements}}, \bibinfo {author}
  {\bibfnamefont {C.~S.}\ \bibnamefont {Kengle}}, \bibinfo {author}
  {\bibfnamefont {S.}~\bibnamefont {Sun}}, \bibinfo {author} {\bibfnamefont
  {K.}~\bibnamefont {Allen}}, \bibinfo {author} {\bibfnamefont
  {Q.}~\bibnamefont {Li}}, \bibinfo {author} {\bibfnamefont {Y.~Y.}\
  \bibnamefont {Peng}}, \bibinfo {author} {\bibfnamefont {A.~A.}\ \bibnamefont
  {Husain}}, \bibinfo {author} {\bibfnamefont {M.}~\bibnamefont {Mitrano}},
  \bibinfo {author} {\bibfnamefont {M.~J.}\ \bibnamefont {Krogstad}}, \bibinfo
  {author} {\bibfnamefont {R.}~\bibnamefont {Osborn}}, \bibinfo {author}
  {\bibfnamefont {A.~B.}\ \bibnamefont {Puthirath}}, \bibinfo {author}
  {\bibfnamefont {S.}~\bibnamefont {Chi}}, \bibinfo {author} {\bibfnamefont
  {L.}~\bibnamefont {Debeer-Schmitt}}, \bibinfo {author} {\bibfnamefont
  {J.}~\bibnamefont {Gaudet}}, \bibinfo {author} {\bibfnamefont
  {P.}~\bibnamefont {Abbamonte}}, \bibinfo {author} {\bibfnamefont {J.~W.}\
  \bibnamefont {Lynn}},\ and\ \bibinfo {author} {\bibfnamefont
  {E.}~\bibnamefont {Morosan}},\ }\bibfield  {title} {\bibinfo {title}
  {Incommensurate magnetic orders and topological hall effect in the square-net
  centrosymmetric \ce{EuGa2Al2} system},\ }\href
  {https://doi.org/10.1103/PhysRevMaterials.6.074201} {\bibfield  {journal}
  {\bibinfo  {journal} {Phys. Rev. Mater.}\ }\textbf {\bibinfo {volume} {6}},\
  \bibinfo {pages} {074201} (\bibinfo {year} {2022})}\BibitemShut {NoStop}%
\bibitem [{\citenamefont {Khanh}\ \emph {et~al.}(2020)\citenamefont {Khanh},
  \citenamefont {Nakajima}, \citenamefont {Yu}, \citenamefont {Gao},
  \citenamefont {Shibata}, \citenamefont {Hirschberger}, \citenamefont
  {Yamasaki}, \citenamefont {Sagayama}, \citenamefont {Nakao}, \citenamefont
  {Peng} \emph {et~al.}}]{khanh2020nanometric}%
  \BibitemOpen
  \bibfield  {author} {\bibinfo {author} {\bibfnamefont {N.~D.}\ \bibnamefont
  {Khanh}}, \bibinfo {author} {\bibfnamefont {T.}~\bibnamefont {Nakajima}},
  \bibinfo {author} {\bibfnamefont {X.}~\bibnamefont {Yu}}, \bibinfo {author}
  {\bibfnamefont {S.}~\bibnamefont {Gao}}, \bibinfo {author} {\bibfnamefont
  {K.}~\bibnamefont {Shibata}}, \bibinfo {author} {\bibfnamefont
  {M.}~\bibnamefont {Hirschberger}}, \bibinfo {author} {\bibfnamefont
  {Y.}~\bibnamefont {Yamasaki}}, \bibinfo {author} {\bibfnamefont
  {H.}~\bibnamefont {Sagayama}}, \bibinfo {author} {\bibfnamefont
  {H.}~\bibnamefont {Nakao}}, \bibinfo {author} {\bibfnamefont
  {L.}~\bibnamefont {Peng}}, \emph {et~al.},\ }\bibfield  {title} {\bibinfo
  {title} {Nanometric square skyrmion lattice in a centrosymmetric tetragonal
  magnet},\ }\href {https://doi.org/https://doi.org/10.1038/s41565-020-0684-7}
  {\bibfield  {journal} {\bibinfo  {journal} {Nat. Nanotechnol.}\ }\textbf
  {\bibinfo {volume} {15}},\ \bibinfo {pages} {444} (\bibinfo {year}
  {2020})}\BibitemShut {NoStop}%
\bibitem [{\citenamefont {Yasui}\ \emph {et~al.}(2020)\citenamefont {Yasui},
  \citenamefont {Butler}, \citenamefont {Khanh}, \citenamefont {Hayami},
  \citenamefont {Nomoto}, \citenamefont {Hanaguri}, \citenamefont {Motome},
  \citenamefont {Arita}, \citenamefont {Arima}, \citenamefont {Tokura} \emph
  {et~al.}}]{yasui2020imaging}%
  \BibitemOpen
  \bibfield  {author} {\bibinfo {author} {\bibfnamefont {Y.}~\bibnamefont
  {Yasui}}, \bibinfo {author} {\bibfnamefont {C.~J.}\ \bibnamefont {Butler}},
  \bibinfo {author} {\bibfnamefont {N.~D.}\ \bibnamefont {Khanh}}, \bibinfo
  {author} {\bibfnamefont {S.}~\bibnamefont {Hayami}}, \bibinfo {author}
  {\bibfnamefont {T.}~\bibnamefont {Nomoto}}, \bibinfo {author} {\bibfnamefont
  {T.}~\bibnamefont {Hanaguri}}, \bibinfo {author} {\bibfnamefont
  {Y.}~\bibnamefont {Motome}}, \bibinfo {author} {\bibfnamefont
  {R.}~\bibnamefont {Arita}}, \bibinfo {author} {\bibfnamefont {T.-H.}\
  \bibnamefont {Arima}}, \bibinfo {author} {\bibfnamefont {Y.}~\bibnamefont
  {Tokura}}, \emph {et~al.},\ }\bibfield  {title} {\bibinfo {title} {Imaging
  the coupling between itinerant electrons and localised moments in the
  centrosymmetric skyrmion magnet {G}d{R}u{$_2$S}i{$_2$}},\ }\href
  {https://doi.org/https://doi.org/10.1038/s41467-020-19751-4} {\bibfield
  {journal} {\bibinfo  {journal} {Nat. Commun.}\ }\textbf {\bibinfo {volume}
  {11}},\ \bibinfo {pages} {5925} (\bibinfo {year} {2020})}\BibitemShut
  {NoStop}%
\bibitem [{\citenamefont {Khanh}\ \emph {et~al.}(2022)\citenamefont {Khanh},
  \citenamefont {Nakajima}, \citenamefont {Hayami}, \citenamefont {Gao},
  \citenamefont {Yamasaki}, \citenamefont {Sagayama}, \citenamefont {Nakao},
  \citenamefont {Takagi}, \citenamefont {Motome}, \citenamefont {Tokura},
  \citenamefont {Arima},\ and\ \citenamefont {Seki}}]{khanh2022zoology}%
  \BibitemOpen
  \bibfield  {author} {\bibinfo {author} {\bibfnamefont {N.~D.}\ \bibnamefont
  {Khanh}}, \bibinfo {author} {\bibfnamefont {T.}~\bibnamefont {Nakajima}},
  \bibinfo {author} {\bibfnamefont {S.}~\bibnamefont {Hayami}}, \bibinfo
  {author} {\bibfnamefont {S.}~\bibnamefont {Gao}}, \bibinfo {author}
  {\bibfnamefont {Y.}~\bibnamefont {Yamasaki}}, \bibinfo {author}
  {\bibfnamefont {H.}~\bibnamefont {Sagayama}}, \bibinfo {author}
  {\bibfnamefont {H.}~\bibnamefont {Nakao}}, \bibinfo {author} {\bibfnamefont
  {R.}~\bibnamefont {Takagi}}, \bibinfo {author} {\bibfnamefont
  {Y.}~\bibnamefont {Motome}}, \bibinfo {author} {\bibfnamefont
  {Y.}~\bibnamefont {Tokura}}, \bibinfo {author} {\bibfnamefont {T.-H.}\
  \bibnamefont {Arima}},\ and\ \bibinfo {author} {\bibfnamefont
  {S.}~\bibnamefont {Seki}},\ }\bibfield  {title} {\bibinfo {title} {Zoology of
  multiple-{Q} spin textures in a centrosymmetric tetragonal magnet with
  itinerant electrons},\ }\href
  {https://doi.org/https://doi.org/10.1002/advs.202105452} {\bibfield
  {journal} {\bibinfo  {journal} {Adv. Sci.}\ }\textbf {\bibinfo {volume}
  {9}},\ \bibinfo {pages} {2105452} (\bibinfo {year} {2022})}\BibitemShut
  {NoStop}%
\bibitem [{\citenamefont {Hou}\ \emph {et~al.}(2021)\citenamefont {Hou},
  \citenamefont {Li}, \citenamefont {Liu}, \citenamefont {Gao}, \citenamefont
  {Ma}, \citenamefont {Zhou}, \citenamefont {Peng}, \citenamefont {Yan},
  \citenamefont {Zhang},\ and\ \citenamefont {Liu}}]{hou2021emergence}%
  \BibitemOpen
  \bibfield  {author} {\bibinfo {author} {\bibfnamefont {Z.}~\bibnamefont
  {Hou}}, \bibinfo {author} {\bibfnamefont {L.}~\bibnamefont {Li}}, \bibinfo
  {author} {\bibfnamefont {C.}~\bibnamefont {Liu}}, \bibinfo {author}
  {\bibfnamefont {X.}~\bibnamefont {Gao}}, \bibinfo {author} {\bibfnamefont
  {Z.}~\bibnamefont {Ma}}, \bibinfo {author} {\bibfnamefont {G.}~\bibnamefont
  {Zhou}}, \bibinfo {author} {\bibfnamefont {Y.}~\bibnamefont {Peng}}, \bibinfo
  {author} {\bibfnamefont {M.}~\bibnamefont {Yan}}, \bibinfo {author}
  {\bibfnamefont {X.-X.}\ \bibnamefont {Zhang}},\ and\ \bibinfo {author}
  {\bibfnamefont {J.}~\bibnamefont {Liu}},\ }\bibfield  {title} {\bibinfo
  {title} {Emergence of room temperature stable skyrmionic bubbles in the rare
  earth based {REM}n{$_2$G}e{$_2$} ({RE~=~C}e, {P}r, and {N}d) magnets},\
  }\href {https://doi.org/https://doi.org/10.1016/j.mtphys.2021.100341}
  {\bibfield  {journal} {\bibinfo  {journal} {Mater. Today Phys.}\ }\textbf
  {\bibinfo {volume} {17}},\ \bibinfo {pages} {100341} (\bibinfo {year}
  {2021})}\BibitemShut {NoStop}%
\bibitem [{\citenamefont {Bouaziz}\ \emph {et~al.}(2022)\citenamefont
  {Bouaziz}, \citenamefont {Mendive-Tapia}, \citenamefont {Bl{\"u}gel},\ and\
  \citenamefont {Staunton}}]{bouaziz2022fermi}%
  \BibitemOpen
  \bibfield  {author} {\bibinfo {author} {\bibfnamefont {J.}~\bibnamefont
  {Bouaziz}}, \bibinfo {author} {\bibfnamefont {E.}~\bibnamefont
  {Mendive-Tapia}}, \bibinfo {author} {\bibfnamefont {S.}~\bibnamefont
  {Bl{\"u}gel}},\ and\ \bibinfo {author} {\bibfnamefont {J.~B.}\ \bibnamefont
  {Staunton}},\ }\bibfield  {title} {\bibinfo {title} {Fermi-surface origin of
  skyrmion lattices in centrosymmetric rare-earth intermetallics},\ }\href
  {https://doi.org/https://doi.org/10.1103/PhysRevLett.128.157206} {\bibfield
  {journal} {\bibinfo  {journal} {Phys. Rev. Lett.}\ }\textbf {\bibinfo
  {volume} {128}},\ \bibinfo {pages} {157206} (\bibinfo {year}
  {2022})}\BibitemShut {NoStop}%
\bibitem [{\citenamefont {Nomoto}\ \emph {et~al.}(2020)\citenamefont {Nomoto},
  \citenamefont {Koretsune},\ and\ \citenamefont
  {Arita}}]{PhysRevLett.125.117204}%
  \BibitemOpen
  \bibfield  {author} {\bibinfo {author} {\bibfnamefont {T.}~\bibnamefont
  {Nomoto}}, \bibinfo {author} {\bibfnamefont {T.}~\bibnamefont {Koretsune}},\
  and\ \bibinfo {author} {\bibfnamefont {R.}~\bibnamefont {Arita}},\ }\bibfield
   {title} {\bibinfo {title} {Formation mechanism of the helical $\mathit{Q}$
  structure in \ce{Gd}-based skyrmion materials},\ }\href
  {https://doi.org/10.1103/PhysRevLett.125.117204} {\bibfield  {journal}
  {\bibinfo  {journal} {Phys. Rev. Lett.}\ }\textbf {\bibinfo {volume} {125}},\
  \bibinfo {pages} {117204} (\bibinfo {year} {2020})}\BibitemShut {NoStop}%
\bibitem [{\citenamefont {Garnier}\ \emph {et~al.}(1996)\citenamefont
  {Garnier}, \citenamefont {Gignoux}, \citenamefont {Schmitt},\ and\
  \citenamefont {Shigeoka}}]{garnier1996giant}%
  \BibitemOpen
  \bibfield  {author} {\bibinfo {author} {\bibfnamefont {A.}~\bibnamefont
  {Garnier}}, \bibinfo {author} {\bibfnamefont {D.}~\bibnamefont {Gignoux}},
  \bibinfo {author} {\bibfnamefont {D.}~\bibnamefont {Schmitt}},\ and\ \bibinfo
  {author} {\bibfnamefont {T.}~\bibnamefont {Shigeoka}},\ }\bibfield  {title}
  {\bibinfo {title} {Giant magnetic anisotropy in tetragonal
  {G}d{R}u{$_2$G}e{$_2$} and {G}d{R}u{$_2$S}i{$_2$}},\ }\href
  {https://doi.org/https://doi.org/10.1016/0921-4526(96)00010-5} {\bibfield
  {journal} {\bibinfo  {journal} {Physica B}\ }\textbf {\bibinfo {volume}
  {222}},\ \bibinfo {pages} {80} (\bibinfo {year} {1996})}\BibitemShut
  {NoStop}%
\bibitem [{\citenamefont {Chapon}\ \emph {et~al.}(2011)\citenamefont {Chapon},
  \citenamefont {Manuel}, \citenamefont {Radaelli}, \citenamefont {Benson},
  \citenamefont {Perrott}, \citenamefont {Ansell}, \citenamefont {Rhodes},
  \citenamefont {Raspino}, \citenamefont {Duxbury}, \citenamefont {Spill} \emph
  {et~al.}}]{chapon2011wish}%
  \BibitemOpen
  \bibfield  {author} {\bibinfo {author} {\bibfnamefont {L.~C.}\ \bibnamefont
  {Chapon}}, \bibinfo {author} {\bibfnamefont {P.}~\bibnamefont {Manuel}},
  \bibinfo {author} {\bibfnamefont {P.~G.}\ \bibnamefont {Radaelli}}, \bibinfo
  {author} {\bibfnamefont {C.}~\bibnamefont {Benson}}, \bibinfo {author}
  {\bibfnamefont {L.}~\bibnamefont {Perrott}}, \bibinfo {author} {\bibfnamefont
  {S.}~\bibnamefont {Ansell}}, \bibinfo {author} {\bibfnamefont {N.~J.}\
  \bibnamefont {Rhodes}}, \bibinfo {author} {\bibfnamefont {D.}~\bibnamefont
  {Raspino}}, \bibinfo {author} {\bibfnamefont {D.}~\bibnamefont {Duxbury}},
  \bibinfo {author} {\bibfnamefont {E.}~\bibnamefont {Spill}}, \emph {et~al.},\
  }\bibfield  {title} {\bibinfo {title} {{WISH}: The new powder and single
  crystal magnetic diffractometer on the second target station},\ }\href
  {https://doi.org/https://doi.org/10.1080/10448632.2011.569650} {\bibfield
  {journal} {\bibinfo  {journal} {Neutron News}\ }\textbf {\bibinfo {volume}
  {22}},\ \bibinfo {pages} {22} (\bibinfo {year} {2011})}\BibitemShut {NoStop}%
\bibitem [{SI()}]{SI}%
  \BibitemOpen
  \href@noop {} {}\bibinfo {note} {See Supplemental Material at [link to
  Supplemental] for S1: Schematic of how the two magnetic domains contribute to
  the diffraction plane. S2: Single crystal neutron diffraction data in the
  $(h,k,0)$ plane showing satellites surrounding the $(-1,1,0)$ nuclear
  peak.}\BibitemShut {Stop}%
\bibitem [{\citenamefont {Izyumov}\ and\ \citenamefont
  {Syromyatnikov}()}]{izyumov1990phase}%
  \BibitemOpen
  \bibfield  {author} {\bibinfo {author} {\bibfnamefont {Y.~A.}\ \bibnamefont
  {Izyumov}}\ and\ \bibinfo {author} {\bibfnamefont {V.~N.}\ \bibnamefont
  {Syromyatnikov}},\ }\href {https://doi.org/10.1007/978-94-009-1920-4} {\emph
  {\bibinfo {title} {Phase transitions and crystal symmetry}}}\ (\bibinfo
  {publisher} {Kluwer Academic Publishers 1990})\BibitemShut {NoStop}%
\bibitem [{\citenamefont {Fawcett}(1988)}]{fawcett1988spin}%
  \BibitemOpen
  \bibfield  {author} {\bibinfo {author} {\bibfnamefont {E.}~\bibnamefont
  {Fawcett}},\ }\bibfield  {title} {\bibinfo {title} {Spin-density-wave
  antiferromagnetism in chromium},\ }\href
  {https://doi.org/10.1103/RevModPhys.60.209} {\bibfield  {journal} {\bibinfo
  {journal} {Rev. Mod. Phys.}\ }\textbf {\bibinfo {volume} {60}},\ \bibinfo
  {pages} {209} (\bibinfo {year} {1988})}\BibitemShut {NoStop}%
\bibitem [{\citenamefont {Kimura}\ \emph {et~al.}(2008)\citenamefont {Kimura},
  \citenamefont {Matsuda}, \citenamefont {Masuda}, \citenamefont {Hondo},
  \citenamefont {Kaneko}, \citenamefont {Metoki}, \citenamefont {Hagiwara},
  \citenamefont {Takeuchi}, \citenamefont {Okunishi}, \citenamefont {He},
  \citenamefont {Kindo}, \citenamefont {Taniyama},\ and\ \citenamefont
  {Itoh}}]{PhysRevLett.101.207201}%
  \BibitemOpen
  \bibfield  {author} {\bibinfo {author} {\bibfnamefont {S.}~\bibnamefont
  {Kimura}}, \bibinfo {author} {\bibfnamefont {M.}~\bibnamefont {Matsuda}},
  \bibinfo {author} {\bibfnamefont {T.}~\bibnamefont {Masuda}}, \bibinfo
  {author} {\bibfnamefont {S.}~\bibnamefont {Hondo}}, \bibinfo {author}
  {\bibfnamefont {K.}~\bibnamefont {Kaneko}}, \bibinfo {author} {\bibfnamefont
  {N.}~\bibnamefont {Metoki}}, \bibinfo {author} {\bibfnamefont
  {M.}~\bibnamefont {Hagiwara}}, \bibinfo {author} {\bibfnamefont
  {T.}~\bibnamefont {Takeuchi}}, \bibinfo {author} {\bibfnamefont
  {K.}~\bibnamefont {Okunishi}}, \bibinfo {author} {\bibfnamefont
  {Z.}~\bibnamefont {He}}, \bibinfo {author} {\bibfnamefont {K.}~\bibnamefont
  {Kindo}}, \bibinfo {author} {\bibfnamefont {T.}~\bibnamefont {Taniyama}},\
  and\ \bibinfo {author} {\bibfnamefont {M.}~\bibnamefont {Itoh}},\ }\bibfield
  {title} {\bibinfo {title} {Longitudinal spin density wave order in a
  quasi-1{D} ising-like quantum antiferromagnet},\ }\href
  {https://doi.org/10.1103/PhysRevLett.101.207201} {\bibfield  {journal}
  {\bibinfo  {journal} {Phys. Rev. Lett.}\ }\textbf {\bibinfo {volume} {101}},\
  \bibinfo {pages} {207201} (\bibinfo {year} {2008})}\BibitemShut {NoStop}%
\bibitem [{\citenamefont {Can\'evet}\ \emph {et~al.}(2013)\citenamefont
  {Can\'evet}, \citenamefont {Grenier}, \citenamefont
  {Klanj\ifmmode~\check{s}\else \v{s}\fi{}ek}, \citenamefont {Berthier},
  \citenamefont {Horvati\ifmmode~\acute{c}\else \'{c}\fi{}}, \citenamefont
  {Simonet},\ and\ \citenamefont {Lejay}}]{PhysRevB.87.054408}%
  \BibitemOpen
  \bibfield  {author} {\bibinfo {author} {\bibfnamefont {E.}~\bibnamefont
  {Can\'evet}}, \bibinfo {author} {\bibfnamefont {B.}~\bibnamefont {Grenier}},
  \bibinfo {author} {\bibfnamefont {M.}~\bibnamefont
  {Klanj\ifmmode~\check{s}\else \v{s}\fi{}ek}}, \bibinfo {author}
  {\bibfnamefont {C.}~\bibnamefont {Berthier}}, \bibinfo {author}
  {\bibfnamefont {M.}~\bibnamefont {Horvati\ifmmode~\acute{c}\else
  \'{c}\fi{}}}, \bibinfo {author} {\bibfnamefont {V.}~\bibnamefont {Simonet}},\
  and\ \bibinfo {author} {\bibfnamefont {P.}~\bibnamefont {Lejay}},\ }\bibfield
   {title} {\bibinfo {title} {Field-induced magnetic behavior in
  quasi-one-dimensional ising-like antiferromagnet {B}a{C}o{$_2$V$_2$O$_8$}: A
  single-crystal neutron diffraction study},\ }\href
  {https://doi.org/10.1103/PhysRevB.87.054408} {\bibfield  {journal} {\bibinfo
  {journal} {Phys. Rev. B}\ }\textbf {\bibinfo {volume} {87}},\ \bibinfo
  {pages} {054408} (\bibinfo {year} {2013})}\BibitemShut {NoStop}%
\bibitem [{\citenamefont {Facheris}\ \emph {et~al.}(2022)\citenamefont
  {Facheris}, \citenamefont {Povarov}, \citenamefont {Nabi}, \citenamefont
  {Mazzone}, \citenamefont {Lass}, \citenamefont {Roessli}, \citenamefont
  {Ressouche}, \citenamefont {Yan}, \citenamefont {Gvasaliya},\ and\
  \citenamefont {Zheludev}}]{PhysRevLett.129.087201}%
  \BibitemOpen
  \bibfield  {author} {\bibinfo {author} {\bibfnamefont {L.}~\bibnamefont
  {Facheris}}, \bibinfo {author} {\bibfnamefont {K.~Y.}\ \bibnamefont
  {Povarov}}, \bibinfo {author} {\bibfnamefont {S.~D.}\ \bibnamefont {Nabi}},
  \bibinfo {author} {\bibfnamefont {D.~G.}\ \bibnamefont {Mazzone}}, \bibinfo
  {author} {\bibfnamefont {J.}~\bibnamefont {Lass}}, \bibinfo {author}
  {\bibfnamefont {B.}~\bibnamefont {Roessli}}, \bibinfo {author} {\bibfnamefont
  {E.}~\bibnamefont {Ressouche}}, \bibinfo {author} {\bibfnamefont
  {Z.}~\bibnamefont {Yan}}, \bibinfo {author} {\bibfnamefont {S.}~\bibnamefont
  {Gvasaliya}},\ and\ \bibinfo {author} {\bibfnamefont {A.}~\bibnamefont
  {Zheludev}},\ }\bibfield  {title} {\bibinfo {title} {Spin density wave versus
  fractional magnetization plateau in a triangular antiferromagnet},\ }\href
  {https://doi.org/10.1103/PhysRevLett.129.087201} {\bibfield  {journal}
  {\bibinfo  {journal} {Phys. Rev. Lett.}\ }\textbf {\bibinfo {volume} {129}},\
  \bibinfo {pages} {087201} (\bibinfo {year} {2022})}\BibitemShut {NoStop}%
\bibitem [{\citenamefont
  {Rodr{\'\i}guez-Carvajal}(1993)}]{rodriguez1993recent}%
  \BibitemOpen
  \bibfield  {author} {\bibinfo {author} {\bibfnamefont {J.}~\bibnamefont
  {Rodr{\'\i}guez-Carvajal}},\ }\bibfield  {title} {\bibinfo {title} {Recent
  advances in magnetic structure determination by neutron powder diffraction},\
  }\href {https://doi.org/https://doi.org/10.1016/0921-4526(93)90108-I}
  {\bibfield  {journal} {\bibinfo  {journal} {Physica B}\ }\textbf {\bibinfo
  {volume} {192}},\ \bibinfo {pages} {55} (\bibinfo {year} {1993})}\BibitemShut
  {NoStop}%
\bibitem [{\citenamefont {Hayami}\ and\ \citenamefont
  {Motome}(2021)}]{hayami2021topological}%
  \BibitemOpen
  \bibfield  {author} {\bibinfo {author} {\bibfnamefont {S.}~\bibnamefont
  {Hayami}}\ and\ \bibinfo {author} {\bibfnamefont {Y.}~\bibnamefont
  {Motome}},\ }\bibfield  {title} {\bibinfo {title} {Topological spin crystals
  by itinerant frustration},\ }\href {https://doi.org/10.1088/1361-648X/ac1a30}
  {\bibfield  {journal} {\bibinfo  {journal} {J. Phys.: Condens. Matter}\
  }\textbf {\bibinfo {volume} {33}},\ \bibinfo {pages} {443001} (\bibinfo
  {year} {2021})}\BibitemShut {NoStop}%
\bibitem [{\citenamefont {Hoffmann}\ and\ \citenamefont
  {Bl\"ugel}(2020)}]{Hoffmann2020}%
  \BibitemOpen
  \bibfield  {author} {\bibinfo {author} {\bibfnamefont {M.}~\bibnamefont
  {Hoffmann}}\ and\ \bibinfo {author} {\bibfnamefont {S.}~\bibnamefont
  {Bl\"ugel}},\ }\bibfield  {title} {\bibinfo {title} {Systematic derivation of
  realistic spin models for beyond-{H}eisenberg solids},\ }\href
  {https://doi.org/10.1103/PhysRevB.101.024418} {\bibfield  {journal} {\bibinfo
   {journal} {Phys. Rev. B}\ }\textbf {\bibinfo {volume} {101}},\ \bibinfo
  {pages} {024418} (\bibinfo {year} {2020})}\BibitemShut {NoStop}%
\bibitem [{\citenamefont {Heinze}\ \emph {et~al.}(2011)\citenamefont {Heinze},
  \citenamefont {Von~Bergmann}, \citenamefont {Menzel}, \citenamefont {Brede},
  \citenamefont {Kubetzka}, \citenamefont {Wiesendanger}, \citenamefont
  {Bihlmayer},\ and\ \citenamefont {Bl{\"u}gel}}]{heinze2011spontaneous}%
  \BibitemOpen
  \bibfield  {author} {\bibinfo {author} {\bibfnamefont {S.}~\bibnamefont
  {Heinze}}, \bibinfo {author} {\bibfnamefont {K.}~\bibnamefont
  {Von~Bergmann}}, \bibinfo {author} {\bibfnamefont {M.}~\bibnamefont
  {Menzel}}, \bibinfo {author} {\bibfnamefont {J.}~\bibnamefont {Brede}},
  \bibinfo {author} {\bibfnamefont {A.}~\bibnamefont {Kubetzka}}, \bibinfo
  {author} {\bibfnamefont {R.}~\bibnamefont {Wiesendanger}}, \bibinfo {author}
  {\bibfnamefont {G.}~\bibnamefont {Bihlmayer}},\ and\ \bibinfo {author}
  {\bibfnamefont {S.}~\bibnamefont {Bl{\"u}gel}},\ }\bibfield  {title}
  {\bibinfo {title} {Spontaneous atomic-scale magnetic skyrmion lattice in two
  dimensions},\ }\href {https://doi.org/10.1038/nphys2045} {\bibfield
  {journal} {\bibinfo  {journal} {Nat. Phys.}\ }\textbf {\bibinfo {volume}
  {7}},\ \bibinfo {pages} {713} (\bibinfo {year} {2011})}\BibitemShut {NoStop}%
\bibitem [{\citenamefont {Sharma}\ \emph {et~al.}(2022)\citenamefont {Sharma},
  \citenamefont {Wang},\ and\ \citenamefont {Batista}}]{sharma2022machine}%
  \BibitemOpen
  \bibfield  {author} {\bibinfo {author} {\bibfnamefont {V.}~\bibnamefont
  {Sharma}}, \bibinfo {author} {\bibfnamefont {Z.}~\bibnamefont {Wang}},\ and\
  \bibinfo {author} {\bibfnamefont {C.~D.}\ \bibnamefont {Batista}},\
  }\bibfield  {title} {\bibinfo {title} {Machine learning assisted derivation
  of effective low energy models for metallic magnets},\ }\href
  {https://doi.org/10.48550/arXiv.2212.09796} {\bibfield  {journal} {\bibinfo
  {journal} {arXiv preprint arXiv:2212.09796}\ } (\bibinfo {year}
  {2022})}\BibitemShut {NoStop}%
\bibitem [{\citenamefont {Mendive-Tapia}\ and\ \citenamefont
  {Staunton}(2017)}]{PhysRevLett.118.197202}%
  \BibitemOpen
  \bibfield  {author} {\bibinfo {author} {\bibfnamefont {E.}~\bibnamefont
  {Mendive-Tapia}}\ and\ \bibinfo {author} {\bibfnamefont {J.~B.}\ \bibnamefont
  {Staunton}},\ }\bibfield  {title} {\bibinfo {title} {Theory of magnetic
  ordering in the heavy rare earths: Ab initio electronic origin of pair- and
  four-spin interactions},\ }\href
  {https://doi.org/10.1103/PhysRevLett.118.197202} {\bibfield  {journal}
  {\bibinfo  {journal} {Phys. Rev. Lett.}\ }\textbf {\bibinfo {volume} {118}},\
  \bibinfo {pages} {197202} (\bibinfo {year} {2017})}\BibitemShut {NoStop}%
\bibitem [{\citenamefont {Balakrishnan}\ \emph {et~al.}(2022)\citenamefont
  {Balakrishnan}, \citenamefont {Wood}, \citenamefont {Manuel}, \citenamefont
  {Hall}, \citenamefont {Orlandi}, \citenamefont {Khalyavin}, \citenamefont
  {Mayoh}, \citenamefont {Petrenko},\ and\ \citenamefont {Lees}}]{Gdata1}%
  \BibitemOpen
  \bibfield  {author} {\bibinfo {author} {\bibfnamefont {G.}~\bibnamefont
  {Balakrishnan}}, \bibinfo {author} {\bibfnamefont {G.~D.~A.}\ \bibnamefont
  {Wood}}, \bibinfo {author} {\bibfnamefont {P.}~\bibnamefont {Manuel}},
  \bibinfo {author} {\bibfnamefont {A.~E.}\ \bibnamefont {Hall}}, \bibinfo
  {author} {\bibfnamefont {F.}~\bibnamefont {Orlandi}}, \bibinfo {author}
  {\bibfnamefont {D.~D.}\ \bibnamefont {Khalyavin}}, \bibinfo {author}
  {\bibfnamefont {D.~A.}\ \bibnamefont {Mayoh}}, \bibinfo {author}
  {\bibfnamefont {O.~A.}\ \bibnamefont {Petrenko}},\ and\ \bibinfo {author}
  {\bibfnamefont {M.~R.}\ \bibnamefont {Lees}},\ }\bibfield  {title} {\bibinfo
  {title} {The skyrmion lattice and other non-collinear spin structures in
  \ce{GdRu2Si2}},\ }\bibfield  {journal} {\bibinfo  {journal} {STFC ISIS
  Facility}\ }\href {https://doi.org/10.5286/ISIS.E.RB2210259}
  {10.5286/ISIS.E.RB2210259} (\bibinfo {year} {2022})\BibitemShut {NoStop}%
\end{thebibliography}%

\end{document}